# Active millimeter wave three-dimensional scan real-time imaging mechanism with a line antenna array

Yang Yu[1,2], Lingbo Qiao[2], Yingxin Wang[1,2], and Ziran Zhao[1,2]*

( 1. Department of Engineering Physics, Tsinghua University, Beijing 100084, China; 2. National Engineering Laboratory for Dangerous Articles and Explosives Detection Technologies, Tsinghua University, Beijing 100084, China.)

**Abstract:** Active Millimeter wave (AMMW) imaging is of interest as it has played important roles in wide variety of applications, from nondestructive test to medical diagnosis. Current AMMW imaging systems have a high spatial resolution and can realize three-dimensional (3D) imaging. However, conventional AMMW imaging systems based on the synthetic aperture require either time-consume acquisition or reconstruction. The AMMW imaging systems based on real-aperture are able to real-time imaging but they need a large aperture and a complex two-dimensional (2D) scan structure to get 3D images. Besides, most AMMW imaging systems need the targets keep still and hold a special posture while screening, limiting the throughput. Here, by using beam control techniques and fast post-processing algorithms, we demonstrate the AMMW 3D scan real-time imaging mechanism with a line antenna array, which can realize 3D real-time imaging by a simple one-dimensional (1D) linear moving, simultaneously, with a satisfactory throughput (over 2000 people per-hour, 10 times than the commercial AMMW imaging systems) and a low system cost. First, the original spherical beam lines generated by the linear antenna array are modulated to fan beam lines via a bi-convex cylindrical lens. Then the holographic imaging algorithm is used to primarily focus the echo data of the imaged object. Finally, the defocus blur is corrected rapidly to get high resolution images by deconvolution. Since our method does not need targets to keep still, has a low system cost, can achieve 3D real-time imaging with a satisfactory throughput simultaneously, this work has the potential to serve as a foundation for future short-range AMMW imaging systems, which can be used in a variety of fields such as security inspection, medical diagnosis, etc.

**Key words:** Millimeter wave; linear array; beam control; real-time imaging algorithm

## 1. Introduction

Due to the advantages of safety, high spatial resolution, and good penetrability, millimeter wave (MMW) imaging is receiving particular interest in the fields of security inspection, medical imaging, and industrial applications [1-7]. Current MMW imaging systems can be divided into active and passive MMW imaging systems according to whether it emits MMW.

Active MMW (AMMW) imaging systems have several advantages over passive MMW imaging systems including high contrast and resolution, better signal noise ratio (SNR), and as such is more

**Foundation item:** Supported by the National Natural Science Foundation of China (61731007)
*Corresponding author: E-mail: zhaozr@mail.tsinghua.edu.cn

suitable for detections of concealed objects. Their application to personnel surveillance has been extensively studied over the past decades [2-3,8-15], and will become the main detection way for personnel surveillance in the future.

The crucial factors in AMMW imaging systems include imaging speed, spatial resolution, and system cost. Furthermore, the throughput must be seriously considered when it comes to massive personnel screening. However, current AMMW imaging systems based on the synthetic aperture require either time-consume acquisition [2-3,9-10] or reconstruction [11-12]. For AMMW imaging systems based on real aperture, quasi-optics technology is usually adopted. Although they can achieve real-time imaging, they need a large aperture and a complex two-dimensional (2D) scan structure to get three-dimensional (3D) images, such as in [13-14]. Besides, most AMMW imaging systems require the targets to maintain still and hold a special posture during the imaging process to prevent image blur, which greatly affects the throughput. Although on the move imaging architecture offers a potential way for high throughput, the massive antennas and tremendous calculated amount prevent it realizing real-time imaging and a low system cost [15]. Therefore, we face a challenge that how to make the AMMW imaging systems have both high resolution and real-time capabilities, and at the same time, with a satisfactory throughput and a low system cost.

Beam control techniques have opened new venues for AMMW imaging [16]. By using beam control techniques, the shape and the directionality of beam can be adjusted in a desired manner [17-20], such as Bessel beam [21], vortex beam [20], and beam deflection [19]. With the help of beam control techniques, the scanning speed of scene can be accelerated and the system cost can be reduced. Then, combined with the most advanced imaging algorithm, the amount of calculation can be greatly reduced, making high resolution real-time imaging with a satisfactory throughput and a low system cost possible. Despite some inspiring advances have been made by applying beam control techniques to AMMW imaging system [22-23], these works have all had a large aperture and an insufficient throughput due to the targets need to stop and hold a special posture while screening. And part of these works lacks the capability of real-time imaging owing to the enormous computational effort [23].

In this study, we propose an AMMW 3D scan real-time imaging mechanism with a line antenna array, which has the advantages of both holographic imaging and real-aperture imaging and can achieve high-resolution real-time imaging with a low system cost simultaneously. In addition, the targets can be imaged while moving and then the throughput can be greatly improved.

## 2. Results

*Design of real-time line scan imager*

As shown in Fig.1, the target under screening stands on the conveyor belt and is moving along horizontal direction (x-direction). We use fan-beam lines, which are modulated from the original spherical beam by a bi-convex cylindrical lens, to illuminate the target to realize real-aperture imaging in the narrow side of the fan-beam (x-direction) and holographic imaging in the wide side of the fan-beam (y-direction).

Then, with the post-processing algorithm developed based on the special shape of the fan-beam, the 3D reconstruction is reduced to a quasi-1D problem, which greatly accelerates the reconstruction speed. Meanwhile, unlike the traditional AMMW holographic imaging, there is no need to scan the complete scene before imaging and the mechanism can image while scanning, which is naturally suitable for parallel computing and real-time imaging. Next, considering the actual arrangement of the linear array, a fast defocus blur correction method based on deconvolution has been developed. Without affecting the real-time performance, the defocus blur is corrected and the spatial resolution can be further improved. Finally, we confirm the effectiveness of the proposed imaging mechanism by the experiments performed on a prototype imager. The prototype imager shows a sub-wavelength lateral resolution and the capability of real-time imaging. Moreover, imaging without stopping targets makes the throughput rapidly increase, which is only limited by the speed of conveyor belt. The throughput of the prototype imager can reach 2000~4000 people per-hour and is 10~20 times higher than the state of art commercial AMMW imaging systems [11].

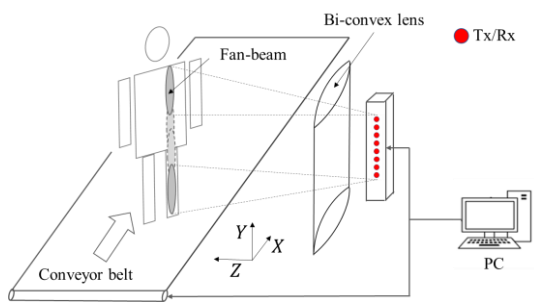

Fig. 1 Schematic diagram of the proposed imaging mechanism.

*Fan-beam modulation*

For the above idea to be implemented, the shape modulation of beam is needed. The common modulation methods of fan-beam include reflector modulation [24], lens modulation [25] and array modulation [26]. Lens modulation has the advantages of large bandwidth, low cost, simple design and convenient feeding. Hence, we select this method as the fan-beam generation method. Teflon is selected as the material of lens for its low dielectric loss and stable performance in millimeter wave band [27].

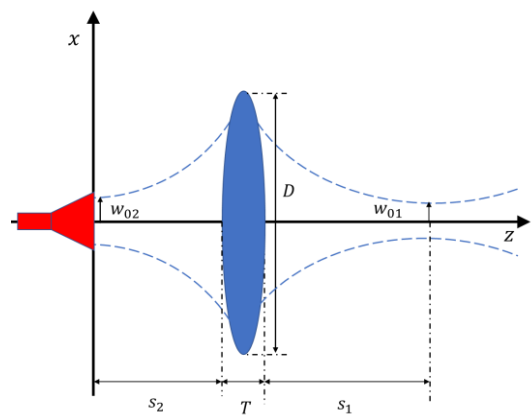

Fig. 2 The illustration of beam modulation with lens

According to Gaussian quasi-optical theory [28], the electric field amplitude distribution of the lens antenna radiation is approximately Gaussian distribution and can be described as:

$$E(x,z) = E_0 \frac{w_0}{w(z)} \exp(-\frac{x^2}{w^2(z)}) \quad (1)$$

where $w_0$ represents the beam radius at the waist position, and $w(z) = w_0\sqrt{1+\left(\frac{\lambda z}{\pi w_0^2}\right)^2}$ is the beam radius at $z$. $x$ is the distance of the field point from the axis.

Derived from Equation (1), the minimum half-power beamwidth (HPBW) is about 1.177 times of $w_0$. The HPBW on object side is related to the spatial resolution, while the HPBW on the image side is limited by the distance between feeds. The effective focal length of lens can be expressed as

$$f = \frac{2(s_1^2 + z_R^2)}{2s_1 + \sqrt{4s_1^2 - 4(s_1^2 + z_R^2)\left(1 - \frac{w_{01}^2}{w_{02}^2}\right)}} \quad (2)$$

where $s_1$ represents the object distance. $z_R = \frac{\pi w_{01}^2}{\lambda}$ is the Rayleigh length. The image distance $s_2$ satisfies the following equation

$$s_2 = f + \frac{s_1 - f}{z_R^2 / f^2 + (s_1 - f)^2} \quad (3)$$

For a hyperboloid lens, the contour curves of the object and image sides can be expressed as

$$z = s_1 + \frac{\sqrt{((n-1)s_1)^2 + (n^2-1)x^2} - (n-1)s_1}{n^2 - 1} \quad (4)$$

$$z = s_1 + T + \frac{\sqrt{((n-1)s_2)^2 + (n^2-1)x^2} - (n-1)s_2}{n^2 - 1} \quad (5)$$

$n$ is the refractive index of the material, $T$ represents the thickness of the lens, and its calculation expression can be written as

$$T = \frac{1}{n+1}\left[\sqrt{s_1^2 + \frac{(n+1)D^2}{4(n-1)}} + \sqrt{s_2^2 + \frac{(n+1)D^2}{4(n-1)}} - (s_1 + s_2)\right] \quad (6)$$

where $D$ represents the aperture size of lens, and it can be derived from Equation (1) that when we set $D = 2.5w$, the interception efficiency of the lens is 95.6 %, which meets the requirements of engineering application.

Then, the effectiveness of generating fan beam with hyperbole cylindrical lens is verified by electromagnetic simulation. Set $s_1$=300 mm $\lambda$=11.11 mm $w_{01}$=8.51 mm, $w_{02}$=17.02 mm, $D$=312.40 mm and the refractive index of Teflon in MMW band is about 1.45. Then we can get $f = 201.40$ mm, $s_2 = 595.79$ mm, $T = 112.12$ mm from Equations (2), (3), (6) and the contour curves of hyperbole cylindrical form (4), (5).

As the length of linear antenna array is about 1000.00 mm, the length of hyperbole cylindrical lens is set as 1600.00 mm.

A pyramidal horn with 60° HPBW on both the E-plane and the H-plane is chosen as the feed antenna. The commercial electromagnetic simulation software FEKO with the Geometrical-Optics (GO) method is used to obtain the results of simulations. The normalized electric filed distribution along z-axis is shown in Fig 3(a). It can be seen from Fig 3(a) that the maximum of electric field occurs at 1008.66 mm, which means that the focal plane locates at this position and it is very close to the theoretical value of 1007.87 mm. Fig.3(b) shows the normalized electric field distribution on focal plane. Seen from simulation result on the focal plane, the HPBW along vertical direction is 650.01 mm, and the HPBW along horizontal direction is 11.35 mm. It means that the original spherical beam emitted by the feed antenna is well transformed into fan-beam. Compared with the theorical value of 10mm, the HPBW along horizontal direction is slightly larger. The main reason is that the radiation of feed is not an ideal Gaussian beam.

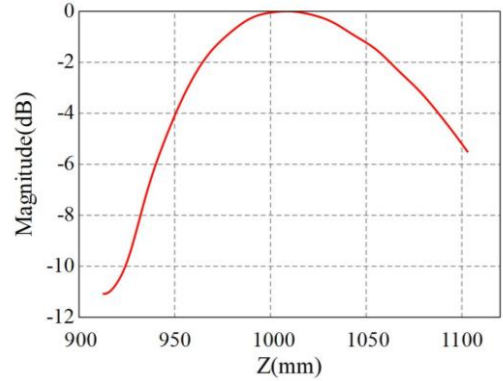

(a)

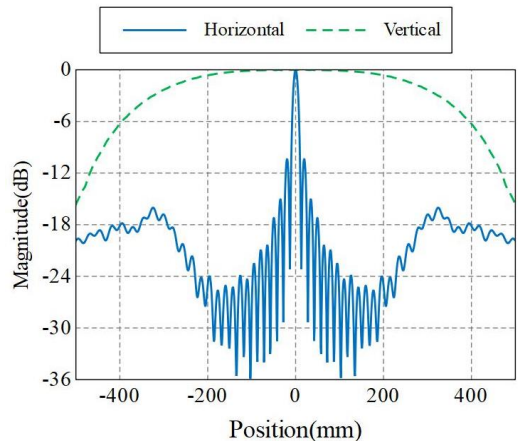

(b)

Fig. 3 Simulation results via FEKO (a) normalized electric

field distribution along z-axis. (b) normalized electric field distribution on focal plane.

*Imaging with fan-beam*

According to quasi-optical theory, the focused beam is well collimated with gradual expansion. Therefore, we can achieve both high resolution in x-direction and the appropriate depth of focusing along z-direction, which is about 200 mm. This is enough for personnel surveillance. Hence, real-aperture imaging can be directly used in x-direction without reconstruction. In the wide-side of fan-beam, holographic method can be adopted to focus images.

The basic imaging configuration is shown in Fig.5. Assuming a general target cell is located at $(x, y, z)$ with a reflectivity of $f(x, y, z)$. Due to the special shape of the fan beam, only when the transceiver is located at the same position with the target cell in x-direction, the echo data will be obtained. Under the Born approximation [29], the scattered filed can be described as:

$$s(x, y_0, k) = \iint f(x, y, z) \frac{e^{-j2kr}}{r^2} dydz \quad (6)$$

where $k = 2\pi/\lambda$ represents the wavenumber of emitting MMW radiation. $r = \sqrt{(y-y_0)^2 + z^2}$ is the distance between the transceiver at $(x, y_0, 0)$ with the target cell at $(x, y, z)$.

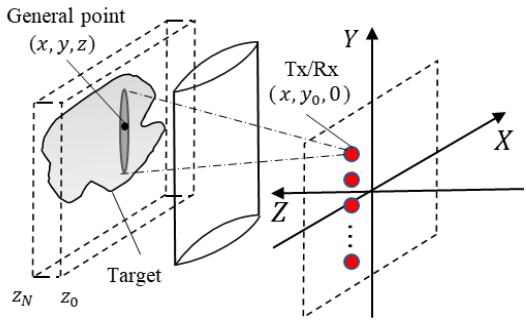

Fig. 5 Illustration of proposed imaging configuration.

According to scalar diffraction theory, the observation field can be written as the superposition of secondary spherical waves scattered from a diffraction aperture. The process can be described by Rayleigh–Sommerfeld diffraction formula

$$U(x, y_0, 0) = \int \frac{1}{j\lambda} U(x, y, z) \frac{e^{jkr}}{r} \cos\theta dy \quad （7）$$

where $U(x, y_0, 0)$ represents the observation field at the plane $z = 0$ and $U(x, y, z)$ is scattered field at the diffraction aperture. $\cos\theta = z/r$ is the inclination factor.

The optical field propagation process is very similar to the AMMW imaging process due to the common superposition calculation [30]. And the measurement of scatter field by AMMW can be regard as an optical field propagation with a wavelength of $\lambda/2$.

Hence, Formula (6) could be further organized as:

$$s^*(x, y_0, k) = \int_{z_0}^{\infty} \frac{j\lambda}{2z} \int \frac{1}{j\lambda/2} f^*(x, y, z) \frac{e^{jk'r}}{r} \cos\theta dydz \quad (8)$$

where $k' = 2k$ and * represents the complex conjugate function, $z_0$ is the front surface of target on Z-axis.

Apart from Rayleigh–Sommerfeld diffraction formula, the angular spectrum formula provides another form to describe optical field propagation process, which uses Fourier transform to analyze the field distribution and can be expressed as

$$U(x, y_0, 0) = \int A(f_y, z) e^{jz\sqrt{4k^2 - k_y^2}} \cdot e^{j2\pi f_y y_0} df_y \quad (9)$$

where $A(f_y, z)$ represents the 1D Fourier transform of $U(x, y, z)$, and $k_y, f_y$ are the spatial wavenumber and the spatial frequency in y-direction, respectively. Since both angular spectrum formula and Rayleigh–Sommerfeld diffraction formula strictly satisfy the

Helmholtz equation and are equivalent to each other, Equation (8) can be rewritten as

$$s^*(x, y_0, k) = \int_{z_0}^{\infty} \frac{j\lambda}{2z} \int A(f_y, z) e^{jz\sqrt{4k^2-k_y^2}} \cdot e^{j2\pi f_y y_0} df_y dz \quad (10)$$

Considering the expression of the Fourier transform, equation (10) can be further rewritten as

$$\begin{aligned} s^*(x, y_0, k) &= \int_0^{+\infty} \frac{j\lambda}{2(z'+z_0)} FT^{-1}_{1D(f_y)}\left[ FT_{1D(y)}\left[ f^*(x,y,z) \right] e^{jk_z(z'+z_0)} \right] dz' \\ &= FT^{-1}_{1D(f_y)}\left[ FT^{-1}_{1D(z')}\left[ \frac{j\lambda}{2(z'+z_0)} FT_{1D(y)}\left[ f^*(x,y,z) \right] \right] e^{jk_z z_0} \right] \\ &= FT^{-1}_{1D(f_y)}\left[ FT_{1D(y)}\left[ FT^{-1}_{1D(z')}\left[ \frac{j\lambda}{2(z'+z_0)} f^*(x,y,z) \right] \right] e^{jk_z z_0} \right] \end{aligned} \quad (11)$$

where $FT^{-1}_{1D(z')}$, $FT^{-1}_{1D(f_y)}$ and $FT_{1D(y)}$ represent 1-D Fourier inverse transform over the variables $z'$, $f_y$ and 1D Fourier transform over $y$, respectively. Thus, the reconstruction algorithm is summarized as

$$f(x,y,z) = \left[ \frac{2z}{j\lambda} e^{-jk_0(z-z_0)} FT_{(k_z)}\left[ FT^{-1}_{1D(k_y)}\left[ stolt\left[ FT_{1D(y_0)}\left[ s^*(x,y_0,k) \right] e^{-jk_z z_0} \right] \right] \right] \right]^* \quad (12)$$

As shown in equation (12), the 3-D image reconstruction is a quasi-1-D problem and the imaging process at each position along x-axis is completely independent. This is a huge advantage for the proposed imaging mechanism. It means that the amount of calculation can be greatly reduced and the proposed imaging mechanism is inherently suitable for parallel computing.

Typically, the scattered filed are sampled discretely to obtain echo data. In order to reconstruct high-quality images, the sampling interval should satisfy the Nyquist criterion. For real aperture imaging in x-direction, the sampling interval should satisfy

$$\Delta x \leq 0.5 \, \delta_{x(\min)} \quad (13)$$

where $\delta_{x(\min)}$ is the minimum resolution in x-direction. In y-direction, the sampling interval should satisfy

$$\Delta y \leq \frac{\lambda_{\min}}{2} \frac{\sqrt{\frac{(L_y+D_y)^2}{4}+Z_0^2}}{L_y+D_y} \quad (14)$$

where $\lambda_{\min}$ represents the shortest wavelength within frequency band. $L_y$, $D_y$ represent the height of array and imaging domain, respectively. However, due to the beam-width of antenna is limited, usually setting the internal as $\lambda_{\min}/2$ is sufficient.

To avoid image aliasing, the frequency sampling interval should satisfy

$$\Delta f \leq \frac{c}{2R_{\max}} \quad (15)$$

where $R_{\max}$ is the maximum of imaging distance.

TABLE I
COMPUTATIONAL COST OF THE PROPOSED ALGORITHMS

| Operation | Real multiplication | Real additions |
|---|---|---|
| 1-D FFT | $2N_f N_y \log_2 N_y$ | $3N_f N_y \log_2 N_y$ |
| Multiplied by $e^{-jk_z z_0}$ | $4N_f N_y$ | $2N_f N_y$ |
| Stolt-interpolation | $6N_f N_y$ | $5N_f N_y$ |
| 1D IFFT | $2N_f N_y \log_2 N_y$ | $3N_f N_y \log_2 N_y$ |
| 1D FFT | $2N_y N_f \log_2 N_f$ | $3N_y N_f \log_2 N_f$ |
| Multiplied by $\frac{2z}{j\lambda}e^{-jk_0(z-z_0)}$ | $4N_z N_y$ | $2N_z N_y$ |
| Conjugation | - | $N_z N_y$ |
| **Total** | $N_y\left[17N_f + 7N_z + 5N_f \log_2\left(N_y^2 N_f\right)\right]$ | |

To quantitatively evaluate the computational complexity of imaging algorithms, we assume that the echo data at each position along x-axis contain $N_y \times N_f$ sampling points. The final computational cost is evaluated by the number of floating-point operations (FLOPs), which can be a real multiplication or a real addition. And the computational cost is given in Table I by the detailed analysis of (12). Considering a classical application for personal screening, the whole scanning scene is about 500 mm × 2000 mm and the typical speed of conveyor belts is 500 mm/s. The frequency band of MMW source is 24-30 GHz and the height of array is 2000 mm.

As mentioned above, the sampling intervals in x-direction, y-direction and frequency are 5.2mm, 5.2mm, and 64MHz, respectively. So the echo data at each position along x-axis consists of $394 \times 88$ sampling point, and the final computational cost is 4.9 MFLOPs. For a PC with GTX 2060 platform, the entire reconstruction required about 0.75 $\mu s$ (considering the GTX 2060 has the computational ability of 6.5 TFLOPS). Due to the time interval between two samples along x-axis is 10.4 ms, it means that the image can be reconstructed in real-time. At the situation, the limitation of throughput is no longer the reconstruction speed or scanning speed but the speed of the conveyor belt.

*Fast correction of defocus blur*

As shown in Fig.6, it is difficult to locate the transmitters and receivers in the linear array at the same position in real applications. Usually, the transmitters and receivers are separated with a small horizontal interval. However, the interval between transmitters and receivers results in serious defocus blur in horizontal direction. To obtain a high-resolution image, the defocus blur must be corrected. At the same time, the correction must be carried out quickly without affecting the real-time performance of AMMW imaging systems. By comprehensive comparison, deconvolution is selected as an effective, robust and fast method to correct the defocus blur.

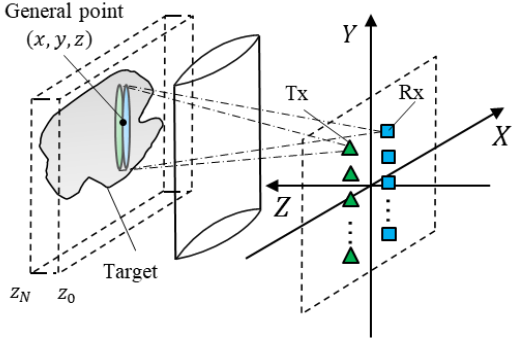

Fig. 6 Illustration of imaging configuration with a separated transmitter and receiver.

Deconvolution is a common method used for reversing the optical distortion in optical images [31-32]. The principle of deconvolution is to assume that the imaging process is that the original image convolves with a point spread function (PSF) and the distorted image is formed by the convolution of original and undistorted image with a distorted PSF. Therefore, we can use deconvolution to reverse the distortion. The key of the method is to obtain the estimation of PSF, and then realize the image calibration.

For the proposed AMMW imaging mechanism, a metal round rod, which is parallel to the line array, can be used as a simulant of the line target, and then we can get the PSF of each channel along x-direction at the same time. Furthermore, by adjusting the distance between the metal rod and the line array, the PSF distribution at different imaging distances is obtained. Since the acquired data is a holographic signal and has been calibrated, there is no need for complex iteration or spectrum estimation. Deconvolution operation can be applied directly and the correction of defocus blur can be described as

$$I(x,y,z) = FT^{-1}_{1D(k_x)} \left[ \left[ FT_{1D(x)} \left[ I_0(x,y,z) \right] / FT_{1D(x)} (PSF) \right] \right]$$

(16)

where $I_0(x,y,z)$ represents the distorted image.

## 3. Application

In order to verify the effectiveness of the proposed imaging mechanism, a prototype imager was constructed. As shown in Fig. 7, the prototype imager consists of four parts: a conveyor belt, a bi-convex cylindrical lens, a line array, and a PC for system controlling and data processing. The bi-convex cylindrical lens has the same parameters as the simulation in section II. The adopted MMW source operates at 24-30GHz range with 64MHz frequency interval. The length of the linear array is 960mm with 5.2 mm sampling interval. The total time it takes for the electrical scanning of linear array to generate a set of echo data is 4.4 ms and it can be reduced further when using more advanced electric switches.

In order to verify the proposed AMMW imaging mechanism, we demonstrate the imaging results of 2D and 3D targets. The photos of those targets are shown in Fig. 8. One is a metal resolution test chart, and the other is a child mannequin that is selected as the 3D target to represent the subjects most at risk. Moreover, detection of a child is harder than an adult for its smaller size. Then those targets are placed 1.20 m distance from the surface of linear array and the speed of the conveyor belt is set as 2km/h.

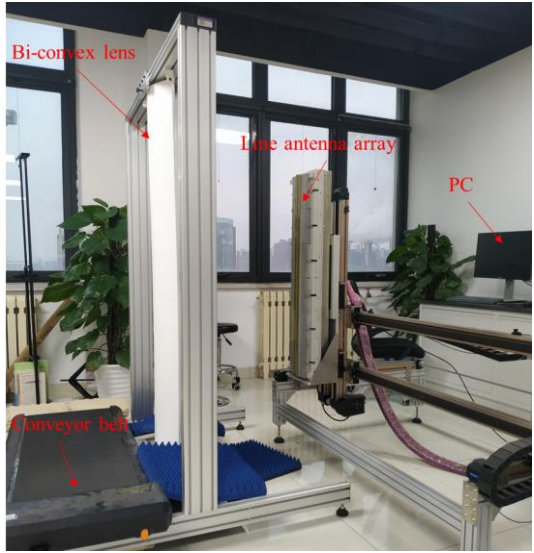

Fig. 7 Photograph of prototype imager.

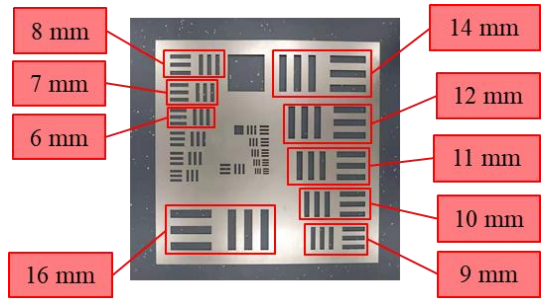

(a)

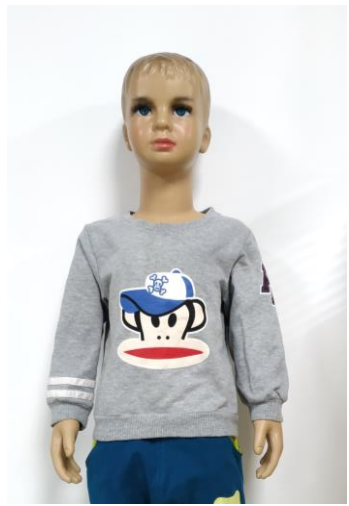

(b)

Fig. 8 Photographs of targets in experiments. (a) resolution test chart (b) child mannequin.

The reconstructed images after the measurements are displayed in Fig. 9 with 18 dB dynamic range. From the imaging result in Fig.9 (a), it could be found that the vertical resolution of the prototype imager can reach 11mm while the horizontal resolution is better than 9 mm. As for the 3D target, the image of the child mannequin is well focused and the details of it can be recognized, which means that the target can be well imaged in the range of 1.1-1.3 m. Furthermore, we can see that the target with a speed of 2 km/h is well imaged, which means the throughput can reach 2000 per hour when assuming that the horizontal distance between two targets is 1 m.

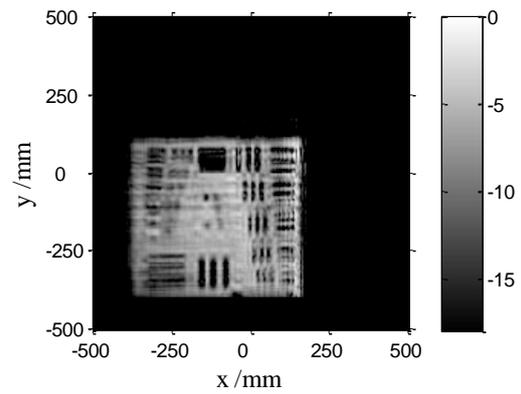

(a)

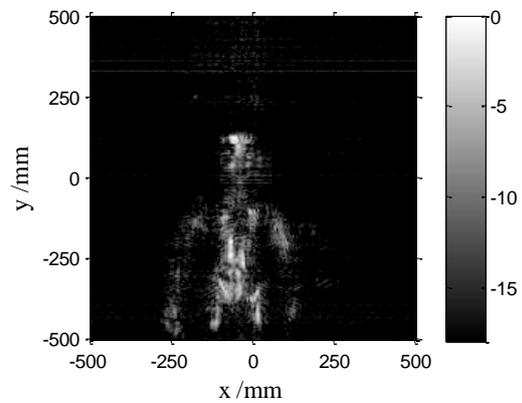

(b)

Fig. 9 Imaging results of the targets with a moving speed of 2km/h. (a) resolution test chart (b) child mannequin.

From all above imaging results, we see that both 2D and 3D targets can be completely focused and accurately imaged by the prototype imager. It demonstrates the effectiveness of the proposed imaging mechanism. In addition, from the imaging result of resolution test chart, it can be seen that the prototype imager has a horizontal resolution better than 9 mm and the vertical resolution can reach 11 mm. Since the wavelength of the adopted MMW source is 10-12 mm, which means that the imaging system has achieved a sub-wavelength lateral resolution.

## 4. Discussion

In this paper, an AMMW 3D scan real-time imaging mechanism with a line antenna array is proposed. The spherical beam is modulated into fan beam by a bi-convex cylindrical lens, which is designed based on quasi-optical theory, and a high-resolution fast imaging algorithm with parallel computing is developed. The horizontal defocus blur existing in the imaging system is corrected by deconvolution method.

The experiments performed on a prototype imager verify the effectiveness of the proposed imaging mechanism. The proposed imaging mechanism can achieve high-resolution real-time imaging with a satisfactory throughput and a low system cost. The unique features of our imaging mechanism are as follows. First, the target is scanned electrically in y-direction. Compared with mechanical scanning, the scanning speed is greatly improved, and the acquisition time of echo data is less than 4.5 microseconds. Secondly, the 3D image reconstruction is transformed as a quasi-1D problem, making the echo data easy to process and the computational complexity is greatly reduced. Thirdly, the 3D image can be captured by a small aperture and a simple 1D linear moving without complex structure, reducing the system cost. Last but the most important is that the imaging process at each position along x-axis is independent, so the imaging system does not need to scan the whole scene before imaging, instead it can image while scanning and the target does not need to keep still.

Future works will concentrate on the improvement of range resolution and the dynamic control of beam via beam control techniques, such as meta-surface and phase arrays. Meanwhile, the efforts will be required to reduce measurement time by employing a MIMO antenna array and more advanced electric switches.

## Reference

[1] Tonouchi, Masayoshi. "Cutting-edge terahertz technology." *Nature photonics* 1.2 (2007): 97-105.

[2] Laviada, Jaime, et al. "Multiview three-dimensional reconstruction by millimetre-wave portable camera." *Scientific reports* 7.1 (2017): 1-11.

[3] Sheen, David M., Douglas L. McMakin, and Thomas E. Hall. "Three-dimensional millimeter-wave imaging for concealed weapon detection." *IEEE Transactions on microwave theory and techniques* 49.9 (2001): 1581-1592.

[4] Stantchev, Rayko Ivanov, et al. "Real-time terahertz imaging with a single-pixel detector." *Nature communications* 11.1 (2020): 1-8.

[5] Rothbart, Nick, et al. "Analysis of human breath by millimeter-wave/terahertz spectroscopy." *Sensors* 19.12 (2019): 2719.

[6] Redo-Sanchez, Albert, et al. "Terahertz time-gated spectral imaging for content extraction through layered structures." *Nature communications* 7.1 (2016): 1-7.

[7] Tong, Jinchao, et al. "Surface plasmon induced direct detection of long wavelength photons." *Nature communications* 8.1 (2017): 1-9.

[8] Li, Lianlin, et al. "Machine-learning reprogrammable metasurface imager." *Nature communications* 10.1 (2019): 1-8.

[9] Sun, Zhaoyang, et al. "Fast three-dimensional image reconstruction of targets under the illumination of terahertz Gaussian beams with enhanced phase-shift migration to improve computation efficiency." *IEEE Transactions on Terahertz Science and Technology* 4.4 (2014): 479-489.

[10] Gao, Jingkun, et al. "Novel efficient 3D short-range imaging algorithms for a scanning 1D-MIMO array." *IEEE Transactions on Image Processing* 27.7 (2018): 3631-3643.

[11] Ahmed, Sherif Sayed, et al. "Advanced microwave imaging." *IEEE microwave magazine* 13.6 (2012): 26-43.

[12] Zhuge, Xiaodong, and Alexander G. Yarovoy. "Study on two-dimensional sparse MIMO UWB arrays for high resolution near-field imaging." *IEEE transactions on antennas and propagation* 60.9 (2012): 4173-4182.

[13] Cooper, Ken B., et al. "THz imaging radar for standoff personnel screening." *IEEE Transactions on Terahertz Science and Technology* 1.1 (2011): 169-182.

[14] Grajal, Jesus, et al. "3-D high-resolution imaging radar at 300 GHz with enhanced FoV." *IEEE Transactions on Microwave Theory and Techniques* 63.3 (2015): 1097-1107.

[15] Gonzalez-Valdes, Borja, et al. "Millimeter wave imaging architecture for on-the-move whole body imaging." *IEEE Transactions on Antennas and Propagation* 64.6 (2016): 2328-2338.

[16] Headland, Daniel, et al. "Tutorial: Terahertz beamforming, from concepts to realizations." *Apl Photonics* 3.5 (2018): 051101.

[17] Walther, Benny, et al. "Spatial and spectral light shaping with metamaterials." *Advanced Materials* 24.47 (2012): 6300-6304.

[18] Hunt, John, et al. "Metamaterial apertures for computational imaging." *Science* 339.6117 (2013): 310-313.

[19] Zhang, Xueqian, et al. "Broadband terahertz wave deflection based on C-shape complex metamaterials with phase discontinuities." *Advanced Materials* 25.33 (2013): 4567-4572.

[20] Genevet, Patrice, et al. "Holographic detection of the orbital angular momentum of light with plasmonic photodiodes." *Nature communications* 3.1 (2012): 1-5.

[21] Shen, Yizhu, et al. "Generating millimeter-wave Bessel beam with orbital angular momentum using reflective-type metasurface inherently integrated with source." *Applied Physics Letters* 112.14 (2018): 141901.

[22] Lyons, Brendan N., Emil Entchev, and Michael K. Crowley. "Reflect-array based mm-wave people screening system." *Millimetre Wave and Terahertz Sensors and Technology VI*. Vol. 8900. International Society for Optics and Photonics, 2013.

[23] Gollub, J. N., et al. "Large metasurface aperture for millimeter wave computational imaging at the human-scale." *Scientific reports* 7.1 (2017): 1-9.

[24] Yurduseven, Okan, and David Smith. "Symmetric/asymmetric H-plane horn fed offset parabolic reflector antenna with switchable pencil/fan-beam radiation characteristics." *ISAPE2012*. IEEE, 2012.

[25] Wu, Xidong, and Jean-Jacques Laurin. "Fan-beam millimeter-wave antenna design based on the cylindrical Luneberg lens." *IEEE Transactions on Antennas and Propagation* 55.8 (2007): 2147-2156.

[26] Falahati, Abolfazl, Mahdi NaghshvarianJahromi, and Robert M. Edwards. "Wideband fan-beam low-sidelobe array antenna using grounded reflector for DECT, 3G, and ultra-wideband wireless applications." *IEEE transactions on antennas and propagation* 61.2 (2012): 700-706.

[27] Bründermann, Erik, Heinz-Wilhelm Hübers, and Maurice FitzGerald Kimmitt. *Terahertz techniques*. Vol. 151. Springer, 2012.

[28] Goldsmith, Paul F. "Quasi-optical techniques at millimeter and submillimeter wavelengths." *Infrared and millimeter waves* 6 (1982): 277-343.

[29] Gubernatis, J. E., et al. "The Born approximation in the theory of the scattering of elastic waves by flaws." *Journal of Applied Physics* 48.7 (1977): 2812-2819.

[30] Qiao, Lingbo, et al. "Exact reconstruction for near-field three-dimensional planar millimeter-wave holographic imaging." *Journal of Infrared, Millimeter, and Terahertz Waves* 36.12 (2015): 1221-1236.

[31] Guo, Min, et al. "Rapid image deconvolution and multiview fusion for optical microscopy." *Nature Biotechnology* 38.11 (2020): 1337-1346.

[32] Pachitariu, Marius, Carsen Stringer, and Kenneth D. Harris. "Robustness of spike deconvolution for neuronal calcium imaging." *Journal of Neuroscience* 38.37 (2018): 7976-7985.